\documentclass[trackchanges,twocolumn]{aastex701}

\begin{document}

\title{Double Power-law Electron Spectra in Solar Flares Due to Temperature Anisotropy Instabilities}

\correspondingauthor{Martín Astete}
%\email{martin.astete@ug.uchile.cl}

%\author{Martín Astete\altaffilmark{1}, 
%        Mario Riquelme\altaffilmark{1}, 
%        Daniel Verscharen\altaffilmark{2}}

%\altaffiltext{1}{Department of Physics, University of Chile, Santiago, Chile}
%\altaffiltext{2}{Mullard Space Science Laboratory, University College London, Dorking, RH5~6NT, United Kingdom}
%\altaffiltext{3}{Another Institution, City, Country}
%\email{martin.astete@ug.uchile.cl, marioriquelme@uchile.cl, d.verscharen@ucl.ac.uk}

\author[orcid=0009-0007-7276-4914]{Martín Astete}
\altaffiliation{martin.astete@ug.uchile.cl}
\affiliation{Department of Physics, University of Chile, Santiago, Chile}
\email{martin.astete@ug.uchile.cl}  

\author[orcid=0000-0003-2928-6412]{Mario Riquelme} 
\altaffiliation{marioriquelme@uchile.cl}
\affiliation{Department of Physics, University of Chile, Santiago, Chile}
\email{marioriquelme@uchile.cl}

\author[orcid=0000-0002-0497-1096]{Daniel Verscharen}
\altaffiliation{d.verscharen@ucl.ac.uk}
\affiliation{Mullard Space Science Laboratory, University College London, Dorking, RH5~6NT, United Kingdom}
\email{d.verscharen@ucl.ac.uk}

\begin{abstract}

Despite extensive observational and theoretical efforts, the physical processes responsible for shaping the diversity of accelerated electron spectra observed in solar flares remain poorly understood. We use 2D particle-in-cell (PIC) simulations of magnetized plasmas subject to continuous shear-driven magnetic amplification to investigate whether electron temperature anisotropy instabilities in above-the-loop-top (ALT) regions can account for this diversity. We explore how the resulting spectra depend on key plasma parameters: the initial electron temperature $T_e$ and the initial ratio of electron cyclotron to plasma frequencies, $f_e = \omega_{ce}/\omega_{pe}$. In our simulations, the adiabatic evolution of the plasma generates electron temperature anisotropy with the electron temperature perpendicular to the magnetic field being larger than the parallel temperature. This eventually drives electromagnetic instabilities capable of scattering and accelerating electrons. The simulations consistently produce nonthermal tails in the electron spectra whose hardness increases with the initial value of $f_e$, while depending only weakly on $T_e$. For runs in which $f_e \lesssim 1.2$, the spectra exhibit double power-law shapes with downward (knee-like) breaks, and the electron scattering is dominated by OQES modes. In runs with $f_e\gtrsim 1.5$, PEMZ modes dominate and produce harder double power-law spectra with upward (elbow-like) breaks. Cases that include the $f_e\sim 1.2-1.5$ transition yield nearly single power-laws that end with bump-like breaks. Our results support the role of temperature anisotropy instabilities in accelerating electrons in ALT regions, offering a promising framework to help explain the wide range of nonthermal electron spectra reported in solar flare observations.

\end{abstract}

\keywords{Sun: flares --- acceleration of particles --- plasmas --- instabilities}


\section{Introduduction} 

Solar flares are some of the most energetic phenomena in the solar system, efficiently converting stored magnetic energy into particle kinetic energy and producing intense radiation across the entire electromagnetic spectrum. A significant fraction of this released energy accelerates electrons to relativistic energies, producing a rich variety of nonthermal electron spectra. However, despite decades of study, the physical processes responsible for this acceleration remain one of the central open questions in heliophysics \cite[see][for a recent review]{2025SSRv..221...27D}.

It is widely accepted that solar flares are triggered by magnetic reconnection in the stressed magnetic fields of the solar corona \citep[e.g.,][]{2020NatAs...4.1140C}. Within the reconnection current sheets (CS), multiple acceleration processes are expected to act concurrently. These include convective and field-aligned electric fields \citep{2005PhRvL..94i5001D,2016ApJ...821...84W} and Fermi-type mechanisms in which electrons gain energy by interacting with merging or contracting magnetic islands \citep{2006Natur.443..553D,2013ApJ...763L...5D,2015ApJ...801..112L,2018ApJ...867...16D,2021PhRvL.126m5101A,2024ApJ...974...47Z}.

Observations also indicate substantial electron acceleration outside the CS, particularly in the above-the-loop-top (ALT) region, located between the bottom of the CS and the apex of the flare loops \citep[e.g.,][]{2020NatAs...4.1140C,2021ApJ...908L..55C,2024ApJ...971...85C}. Here, reconnection outflows can dissipate energy, creating favorable conditions for acceleration. For example, collisions between the CS outflows and loop tops may generate a termination shock \citep{2019ApJ...884...63C,2021ApJ...911....4L}, at which diffusive shock acceleration can occur \citep{2015Sci...350.1238C}. Furthermore, magnetohydrodynamic (MHD) simulations \citep{2022NatAs...6..317S,2023ApJ...943..106S} and observations \citep{2025ApJ...984L..27X,2025ApJ...987..116G,2025ApJ...986L..16F} show that the ALT region becomes strongly turbulent, enabling stochastic electron acceleration by plasma waves \citep{1992ApJ...398..350H,1996ApJ...461..445M,1997ApJ...491..939M,2004ApJ...610..550P,2014ApJ...796...45P}. This picture is supported by recent observations in which space and time correlation between turbulent energy dissipation and electron acceleration are found in ALT regions \citep{2024ApJ...973...96A}.

\citet{2022ApJ...924...52R} (hereafter, Paper~I) use two-dimensional (2D) particle-in-cell (PIC) simulations to demonstrate that plasma waves can also emerge from instabilities driven by electron temperature anisotropies, leading to efficient stochastic acceleration in ALT regions. These anisotropies are expected in weakly collisional turbulent plasmas, in which magnetic field variations from plasma turbulence cause the parallel and perpendicular electron temperatures ($T_{e,\parallel}$ and $T_{e,\perp}$) to diverge through adiabatic evolution \citep{1956RSPSA.236..112C}. Such anisotropies can also arise naturally in ALT regions due to their expected mirror-like magnetic field configurations, which trap electrons with large pitch-angles, forming a loss-cone distribution and enhancing $T_{e,\perp}$ relative to $T_{e,\parallel}$ \citep[e.g.,][]{1998PhyU...41.1157F,2020NatAs...4.1140C}. By driving a continuous growth of the background magnetic field through an externally sustained shear flow, Paper~I  explores the long-term impact of temperature anisotropy instabilities on electron spectra in the case with $T_{e,\perp} > T_{e,\parallel}$, finding that they indeed drive significant acceleration under ALT conditions.

Here, we extend these results by conducting 2D PIC simulations to examine how variations in the initial plasma parameters influence both the efficiency and spectral characteristics of this acceleration. Our main motivation comes from flare observations showing a broad diversity and temporal evolution of nonthermal electron spectra, often with different forms of double power-law distributions. This diversity suggests the coexistence of multiple acceleration processes across different flares or, alternatively, a high sensitivity of one or a few mechanisms to local plasma conditions. In this study, we focus on the latter possibility, exploring how acceleration from temperature anisotropy instabilities in ALT regions responds to changing plasma parameters.

This paper is organized as follows. In \S~\ref{sec:motivation}, we present observational evidence for the diversity of electron spectra in solar flares, which includes double power-law nonthermal electron spectra. In \S~\ref{sec:numsetup}, we describe our shearing setup and simulation configuration. In \S \ref{sec:visheating}, we depict the expected interplay between temperature anisotropies and temperature anisotropy instabilities in our shearing setup, focusing on the characteristics of the expected unstable modes under ALT plasma conditions. In \S \ref{sec:enta}, we show how the time evolution of the nonthermal electron spectra depends on plasma parameters. Finally, in \S \ref{sec:conclusions}, we present our summary and conclusions.

\section{Evidence for Double Power-Law Electron Spectra}
\label{sec:motivation}

Nonthermal hard X-ray (HXR) emission, which originates primarily from the footpoints of flare loops, provides one of the most direct diagnostics of the spectra of accelerated electrons in solar flares. The shapes of these spectra are very diverse and often well described by double (or broken) power laws. They most commonly exhibit a downward, knee-like break \citep{1987ApJ...312..462L,1988ApJ...327..466S,1992ApJ...389..756D,2003ApJ...595L..97H,2019SoPh..294..105A}. While such breaks can sometimes be attributed to instrumental effects (e.g., pulse pileup; \citealt{2002SoPh..210...33S}) or electron transport \citep[e.g.,][]{2011ApJ...731..106S}, they are generally interpreted as intrinsic to the acceleration process \citep[e.g.,][]{2019SoPh..294..105A}. This interpretation is supported by limb-occulted flare observations \citep{2017ApJ...835..124E}, in which the HXR emission originates near the reconnection region and is largely unaffected by propagation effects.

A statistical study of 65 flares by \citet{2019SoPh..294..105A} finds that nearly all HXR spectra can be reproduced by double power-law electron distributions, with spectral breaks occurring near a typical electron energy $\varepsilon_b \sim $100~keV. While downward breaks dominate, some events exhibit a single power law or an upward, elbow-like break. The latter, sometimes observed at energies approaching hundreds of keV, signal intrinsic spectral hardening at high energies \citep{1988SoPh..118...49D,1975SoPh...43..415S,1985JPSJ...54.4462Y}. Supporting this, \citet{2013ApJ...763...87A} report that spectral indices inferred from microwave observations are harder than those derived from HXRs, also suggesting a hardening of the electron spectrum at hundreds of keV energies.

Temporal evolution in spectral shape is also common. \citet{1992ApJ...389..756D} find that during peak HXR emission, most spectra exhibit downward breaks near $\varepsilon_b \sim$100~keV, but in later stages often transition to upward breaks around $\varepsilon_b \sim$70~keV. The limb-occulted GOES X8.2-class flare on 2017 September~10, where the emission originates near the acceleration region, shows similar behavior: during the impulsive phase, \citet{2021ApJ...908L..55C} report a double power-law electron spectrum breaking downward at $\sim$160 keV, whereas near peak HXR emission, \citet{2019RAA....19..173N} identify an elbow-like spectrum.

In situ measurements of solar energetic electrons (SEEs) provide further evidence for double power-law features at the acceleration site. \citet{2023ApJ...948...51W} analyze 458 SEE events observed by \textit{Wind}/3DP over 25 years. Most of these events show clear velocity dispersion, indicative of scatter-free propagation, which may in principle minimize transport effects. Of these, 304 events exhibit downward breaks (typically around $\varepsilon_b \sim$ 60~keV), 32 show upward breaks (near $\varepsilon_b \sim$5~keV), and 23 display single power laws. In addition, \citet{2025A&A...699A...2L} identify 9 events characterized by single power-laws that end with bump-like breaks at tens of keV.

Different acceleration mechanisms have the potential to produce this variety of nonthermal electron spectra under flare conditions. For instance, diffusive shock acceleration at finite-width termination shocks  leads to spectral hardening at high energies \citep{2013ApJ...769...22L,2019ApJ...883...49K,2022ApJ...932...87X}. Likewise, as shown in Paper~I, single power laws with bump-like break features can arise from temperature anisotropy instabilities. However, a significant sensitivity of one or a few mechanisms to variations in local plasma conditions may also contribute to the diversity of spectra. In this work, we consider this possibility with focus on electron acceleration by temperature anisotropy instabilities.

\section{Simulation Strategy and Setup}
\label{sec:numsetup}

To investigate the influence of changes in plasma conditions on stochastic electron acceleration driven by temperature anisotropy instabilities, we vary two key parameters: the initial electron temperature $\Theta_e \equiv T_e/m_ec^2$, where $m_e$ and $c$ are the electron mass and the speed of light, and the initial ratio
\[
f_e \equiv \frac{\omega_{ce}}{\omega_{pe}},
\]
which measures the relative importance of magnetic field strength to plasma density via the cyclotron frequency $\omega_{ce} = eB / m_e c$ and the plasma frequency $\omega_{pe} = \sqrt{4\pi n_e e^2 / m_e}$, where $e$ ($>0$), $B$ and $n_e$ are the elementary charge, the magnetic field magnitude and the electron density.

\subsection{Simulation Strategy}
\label{sec:strategy}

In order to explore the long term effect of temperature anisotropy instabilities on nonthermal electron spectra, as in Paper~I, we generate temperature anisotropy by imposing a macroscopic plasma shear that amplifies the local magnetic field through flux freezing. In a collisionless plasma, this amplification naturally produces $\Theta_{e,\perp} > \Theta_{e,\parallel}$ due to the adiabatic response described by the Chew–Goldberger–Low (CGL) equations of state \citep{1956RSPSA.236..112C}. The anisotropy grows until it crosses a threshold for an instability driven by temperature anisotropy, triggering the rapid growth and saturation of unstable modes. These modes rapidly interact with the electrons, disrupting the CGL evolution. The system then enters a nonlinear stage in which the anisotropy self-regulates to a marginally stable level. This  behavior is consistent with in situ measurements of electron anisotropy in the solar wind \citep[e.g.,][]{2008JGRA..113.3103S}. Our simulations capture this long-term nonlinear regime self-consistently, enabling us to study the sustained effect of these instabilities on electron acceleration.

We focus on the microphysics of the interactions between electrons and the unstable modes by resolving kinetic length scales of order the electron Larmor radius, $R_{Le}$. We model the shear as an external driver representing local fluid velocities from large-scale turbulence. To optimize computational resources, we treat ions as infinitely massive, so their role is limited to providing a neutralizing background charge.

Initial parameters are chosen to reflect typical conditions in ALT regions. Observations indicate electron temperatures of a few tens of MK \citep{1994ApJ...421..843F,1994Natur.371..495M,1995PASJ...47..677M,2011SSRv..159...19F}, so we adopt initial electron temperatures of $T_e^{\mathrm{init}} = 26$ and $52$~MK, corresponding to $\Theta_e^{\mathrm{init}} \equiv k_B T_e^{\mathrm{init}} / m_e c^2 = 0.00438$ and $0.00875$, respectively, where $k_B$ is Boltzmann’s constant. Magnetic field strengths in these regions are typically $\sim100$~G \citep{2019ApJ...874..126K}, with electron densities $n_e \sim 10^9$–$10^{11}$~cm$^{-3}$ \citep{1994ApJ...421..843F,1994Natur.371..495M,1995PASJ...47..677M,1997ApJ...478..787T}, yielding $f_e \sim 0.1$–1. We therefore explore initial values of $f_e$ given by $f_e^{\mathrm{init}} = 0.264$, $0.374$, $0.529$, and $0.748$. These correspond to initial electron plasma beta values $\beta_e^{\mathrm{init}} = 8\pi n_e k_B T_e / B^2$ ranging from 0.0156 to 0.25. We run our simulations until the shear amplifies $B$ by approximately a factor of 4 (assuming constant $n_e$), producing final values of $f_e$ in the range $\sim 1-3$. This approach enables us to systematically assess how $T_e$ and $f_e$ influence both the dominant unstable modes and their acceleration efficiency.

\subsection{Simulation Setup}
\label{sec:setup}

We perform our simulations with the relativistic electromagnetic PIC code TRISTAN-MP \citep{buneman1993tristan,2005AIPC..801..345S}. The computational domain is a square in the $x$–$y$ plane, initially filled with a uniform isotropic Maxwell–Boltzmann electron distribution. We impose an initially uniform magnetic field $\mathbf{B}_0 = B_0 \hat{x}$ which is then amplified by a background shear flow $\mathbf{v} = -s x \, \hat{y}$, where $s$ is the constant shear rate and $x$ is the position along $\hat{x}$ (see Fig.~1 in \citealt{2019ApJ...880..100L}).\footnote{The simulations are conducted in a shearing coordinate frame in which the bulk velocity vanishes, requiring modified forms of Maxwell’s equations and the Lorentz force \citep{2012ApJ...755...50R}.}

By flux conservation, the mean $y$-component of the magnetic field evolves as $\langle B_y \rangle = -s B_0 t$, where angle brackets denote domain-averaged quantities. The total mean field then increases as $|\langle \mathbf{B} \rangle| = B_0 \sqrt{1 + (st)^2}$. This growth conserves the electron magnetic moment $\mu_e$, driving $\Theta_{e,\perp} > \Theta_{e,\parallel}$ until instabilities develop and break $\mu_e$ conservation. We evolve the system until $t\cdot s = 4$, corresponding to a $\sim 4\times$ increase in $|\langle \mathbf{B} \rangle|$.

Another key parameter in our simulations is the time-scale separation between the initial electron cyclotron frequency and $s$, expressed as the ratio $\omega_{ce}^{\mathrm{init}}/s$. Since $s$ represents the inverse of the characteristic time-scale for magnetic field amplification within ALT turbulence, we can estimate it as the inverse of the collapse time of the contracting loop tops into more stable configurations. Given that the outflow velocity from reconnection current sheets is expected to be comparable to the Alfvén speed $v_A$, we approximate $s$ by dividing the typical $v_A$ in the loop tops by their characteristic length scale, $L_{LT}$.  Using our fiducial values $n_e \sim 10^9$--$10^{11}\,\mathrm{cm}^{-3}$, $B \sim 100\,\mathrm{G}$, and $L_{LT} \sim 10^9\,\mathrm{cm}$ \citep[e.g.,][]{2020NatAs...4.1140C}, we obtain $s \sim 0.1$-$1\,\mathrm{s}^{-1}$, which corresponds to $\omega_{ce}/s \sim 10^9$-$10^{10}$. These values of $\omega_{ce}/s$ are several
orders of magnitude larger than what can be achieved in our
simulations. However, in this work we show that, for sufficiently large values of $\omega_{ce}/s$, our results are fairly independent of this parameter (this is checked in \S \ref{sec:realistic}).

Numerical parameters include the number of macro-electrons per cell ($N_{\mathrm{epc}}$), the electron skin depth $d_e = c / \omega_{pe}$ in units of the grid spacing $\Delta_x$ (since the average $n_e$ does not change in our shearing setup, $d_e$ does not evolve), the initial box size $L$ in units of the initial Larmor radius $R_{Le}^{\mathrm{init}} = v_{th,e} / \omega_{ce}^{\mathrm{init}}$ (with $v_{th,e}^2 = k_B T_e^{\mathrm{init}} / m_e$), and the speed of light $c$ in units of $\Delta_x / \Delta_t$, where $\Delta_t$ is the simulation time step. We have verified that our results are robust to variations in these parameters. Table~\ref{tab:param} summarizes our runs.

\begin{table}
\begin{center}
%\begin{threeparttable}

%\tablehead{Simulations parameters}
\begin{tabular}{llcccl}
\hline \hline 
Runs & $N_{\textrm{epc}}$& $d_e/\Delta_x$ & $f_e^{\textrm{init}}$ & $\Theta_e^{\textrm{init}}$ & $\omega_{ce}^{\textrm{init}}/s$\\ \vspace{-0.3cm}\\
\hline 
%\startdata
Fe1Te1-300 & 100 & 15 & 0.264 & 0.00438 & 300 \\ 
Fe1Te1-600 & 100 & 20 & 0.264 & 0.00438 & 600 \\
Fe1Te1-1200 & 200 & 15 & 0.264 & 0.00438 & 1200 \\ 
\hline
Fe2Te1-300 & 100 & 30 & 0.374 & 0.00438 & 300 \\
Fe2Te1-600 & 100 & 30 & 0.374 & 0.00438 & 600 \\
Fe2Te1-1200 & 100 & 30 & 0.374 & 0.00438 & 1200 \\
\hline
Fe3Te1-300 & 100 & 25 & 0.529 & 0.00438 & 300 \\ 
Fe3Te1-600 & 100 & 25 & 0.529 & 0.00438 & 600 \\
Fe3Te1-1200 & 200 & 25 & 0.529 & 0.00438 & 1200 \\ 
\hline
Fe4Te1-1200 & 100 & 30 & 0.748 & 0.00438 & 1200 \\
Fe4Te1-2400 & 100 & 30 & 0.748 & 0.00438 & 2400 \\
Fe4Te1-4800 & 100 & 30 & 0.748 & 0.00438 & 4800 \\
\hline
\hline
Fe1Te2-300 & 100 & 15 & 0.264 & 0.00875 & 300 \\ 
Fe1Te2-600 & 200 & 15 & 0.264 & 0.00875 & 600 \\
Fe1Te2-1200 & 400 & 15 & 0.264 & 0.00875 & 1200 \\ 
\hline
Fe2Te2-300 & 100 & 30 & 0.374 & 0.00875 & 300 \\ 
Fe2Te2-600 & 100 & 30 & 0.374 & 0.00875 & 600 \\ 
Fe2Te2-1200 & 100 & 20 & 0.374 & 0.00875 & 1200 \\ 
\hline
Fe3Te2-300 &  100 & 25 & 0.529 & 0.00875 & 300 \\ 
Fe3Te2-600 &  100 & 25 & 0.529 & 0.00875 & 600 \\
Fe3Te2-1200 & 100 & 25 & 0.529 & 0.00875 & 1200 \\ 
\hline
Fe4Te2-600 & 100 & 25 & 0.748 & 0.00875 & 600 \\
Fe4Te2-1200 & 200 & 25 & 0.748 & 0.00875 & 1200 \\
Fe4Te2-2400 & 200 & 30 & 0.748 & 0.00875 & 2400 \\
\hline
\hline
\vspace{-0.3cm}
\end{tabular}
\caption{Simulation parameters for the presented runs. These are the number of macro-electrons per cell, N$_{\textrm{epc}}$, the electron skin depth, $d_e$, normalized by the grid point spacing $\Delta_x$, the initial ratio between the electron cyclotron and plasma frequencies, $f_e^{\textrm{init}}$ ($\equiv \omega_{c,e}^{\textrm{init}}/\omega_{p,e}$), the initial electron temperature, $\Theta_e^{\textrm{init}}$ ($\equiv T_e^{\textrm{init}}/m_ec^2$), and the ratio between the initial cyclotron frequency and $s$, $\omega_{c,e}^{\textrm{init}}/s$. In all the presented simulations, we use the same values for the box size in terms of the initial electron Larmor radius, $L/R_{Le}^{\textrm{init}}=140$ and the same speed of light value ($c/[\Delta_x/\Delta_t]=0.225$), where $\Delta_t$ is the simulation time step. \label{tab:param}}
%  \end{threeparttable}
  \end{center}
\end{table}

\section{Evolution of electron temperature anisotropy}
\label{sec:visheating}

\noindent Before describing the effects of initial plasma conditions on electron acceleration by temperature anisotropy instabilities, we briefly describe the expected interplay between temperature anisotropies and temperature anisotropy instabilities in our shearing setup. We give special attention to the characteristics of the expected unstable modes under ALT conditions. For this, we use simulation Fe3Te2-1200 (see Table~\ref{tab:param}), which is the same run analyzed in detail in Paper ~I.
\begin{figure*}[t!]  
\centering 
\hspace*{-.2cm}\includegraphics[width=18.4cm]{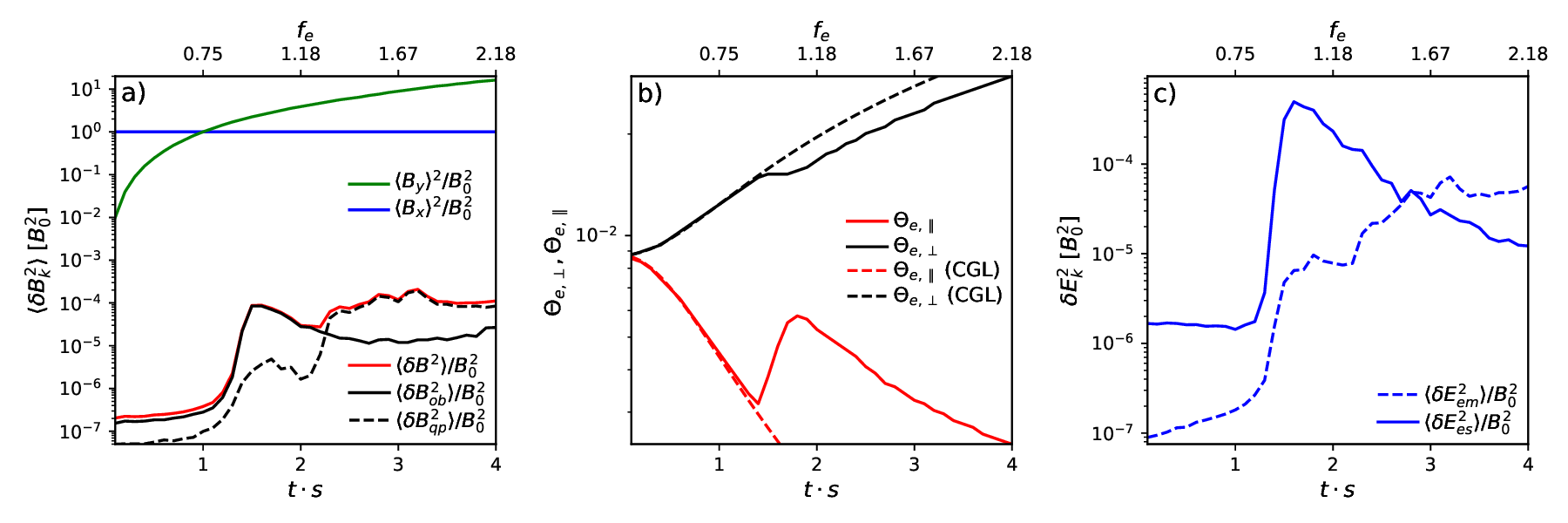}
\caption{Fields and electron temperatures for run Fe3Te2-1200 as functions of normalized time $t\cdot s$ (lower horizontal axes) and of the instantaneous $f_e$ (upper horizontal axes; using the average magnetic field at each time). Panel $\it{a}$ shows in solid-blue and solid-green lines the evolution of the energy in the $x$  and $y$ components of the mean magnetic field $\langle \textbf{\textit{B}}\rangle$, as well as the energy in $\delta \textbf{\textit{B}}$ in solid-red line. The solid-black and dashed-black lines show the contributions to the $\delta \textbf{\textit{B}}$ energy given by the oblique ($\delta \textbf{\textit{B}}_{ob}$) and quasi-parallel ($\delta \textbf{\textit{B}}_{qp}$) modes. Panel $\it{b}$ shows in solid-black (solid-red) the evolution of the electron temperature perpendicular (parallel) to $\langle \textbf{\textit{B}}\rangle$. The dashed-black (dashed-red) line shows the CGL prediction for the perpendicular (parallel) temperature. Finally, panel $\it{c}$ shows in solid- and dashed-blue lines the contributions to the  energy in the fluctuating electric field $\delta \textbf{\textit{E}}$ given by its electrostatic ($\delta \textbf{\textit{E}}_{es}$) and electromagnetic ($\delta \textbf{\textit{E}}_{em}$) components, respectively [all fields energies are in units of the initial magnetic energy]. This figure is adapted from Fig. 1 of Paper I.} 
\label{fig:mubreak} 
\end{figure*}

\subsection{Magnetic Field Amplification and Its Impact on Temperature Anisotropy}
\label{interplay}
Figure~\ref{fig:mubreak}$a$ shows in solid green the linear amplification of the $y$-component of the mean magnetic field $\langle \textbf{\textit{B}} \rangle$ for run Fe3Te2-1200. As expected for our shearing configuration, the $x$-component of the field remains constant at $B_0$, while the $z$-component (perpendicular to the simulation plane) remains at a value of zero. As shown in Figure~\ref{fig:mubreak}$b$,  the increase in $|\langle \textbf{\textit{B}} \rangle|$ produces a corresponding increase in the perpendicular electron temperature $\Theta_{e,\perp}$ (shown in solid black) and a decrease in the parallel temperature $\Theta_{e,\parallel}$ (shown in solid red). Initially, this behavior matches the adiabatic response expected for a collisionless plasma, as described by the CGL equations of state, shown by the dashed black and red lines.\footnote{In the CGL model, $T_{e,\perp}/B$ and $T_{e,\parallel} B^2/n_e^2$ are conserved.} A clear departure from CGL predictions occurs at $t \cdot s \approx 1.4$, which coincides with a rapid growth and saturation of the fluctuating magnetic field $\delta \textbf{\textit{B}}$ ($\equiv \textbf{\textit{B}} - \langle \textbf{\textit{B}} \rangle$), as shown by the solid red curve in Figure~\ref{fig:mubreak}$a$. This time marks the point at which temperature anisotropy is curtailed by the onset of temperature anisotropy-driven instabilities, which break the adiabatic evolution. After $t \cdot s \approx 1.4$, the amplitude of $\delta \textbf{\textit{B}}$ reaches a quasi-steady state, as shown by the red line in Figure~\ref{fig:mubreak}$a$. The subsequent evolution of $\Theta_{e,\perp}$ and $\Theta_{e,\parallel}$ suggests a competition between pitch-angle scattering, which reduces anisotropy, and background field amplification, which enhances it.  $\Theta_{e,\perp} - \Theta_{e,\parallel}$ first decreases rapidly between $t \cdot s \approx 1.4$ and $2$, and then continues to grow, but at a lower rate than predicted by the CGL equations. This behavior signals the transition to a different scattering regime, controlled by the dominance of distinct unstable modes, as we explain below. 
\begin{figure}[t!]  
\centering 
\hspace*{-.4cm}\includegraphics[width=9.3cm]{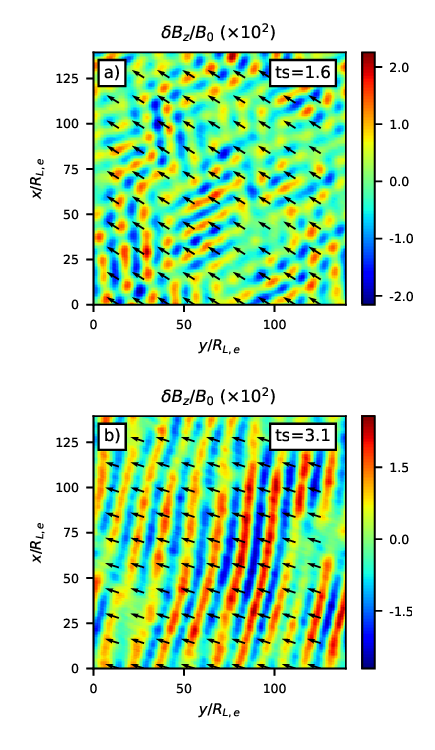}
\caption{The 2D structure of $\delta \textit{B}_z$ at $t\cdot s = 1.6$ (panel $a$) and $t\cdot s = 3.1$ (panel $b$) for run Fe3Te2-1200. $\delta \textit{B}_z$ is normalized by $B_0$. The black arrows show the direction of the average magnetic field $\langle \textbf{\textit{B}} \rangle$.} 
\label{fig:2dstructure} 
\end{figure}

\subsection{The nature of the unstable modes}
\label{sec:nature}
\begin{figure*}[t!]  
\centering 
\hspace*{-.2cm}\includegraphics[width=18.5cm]{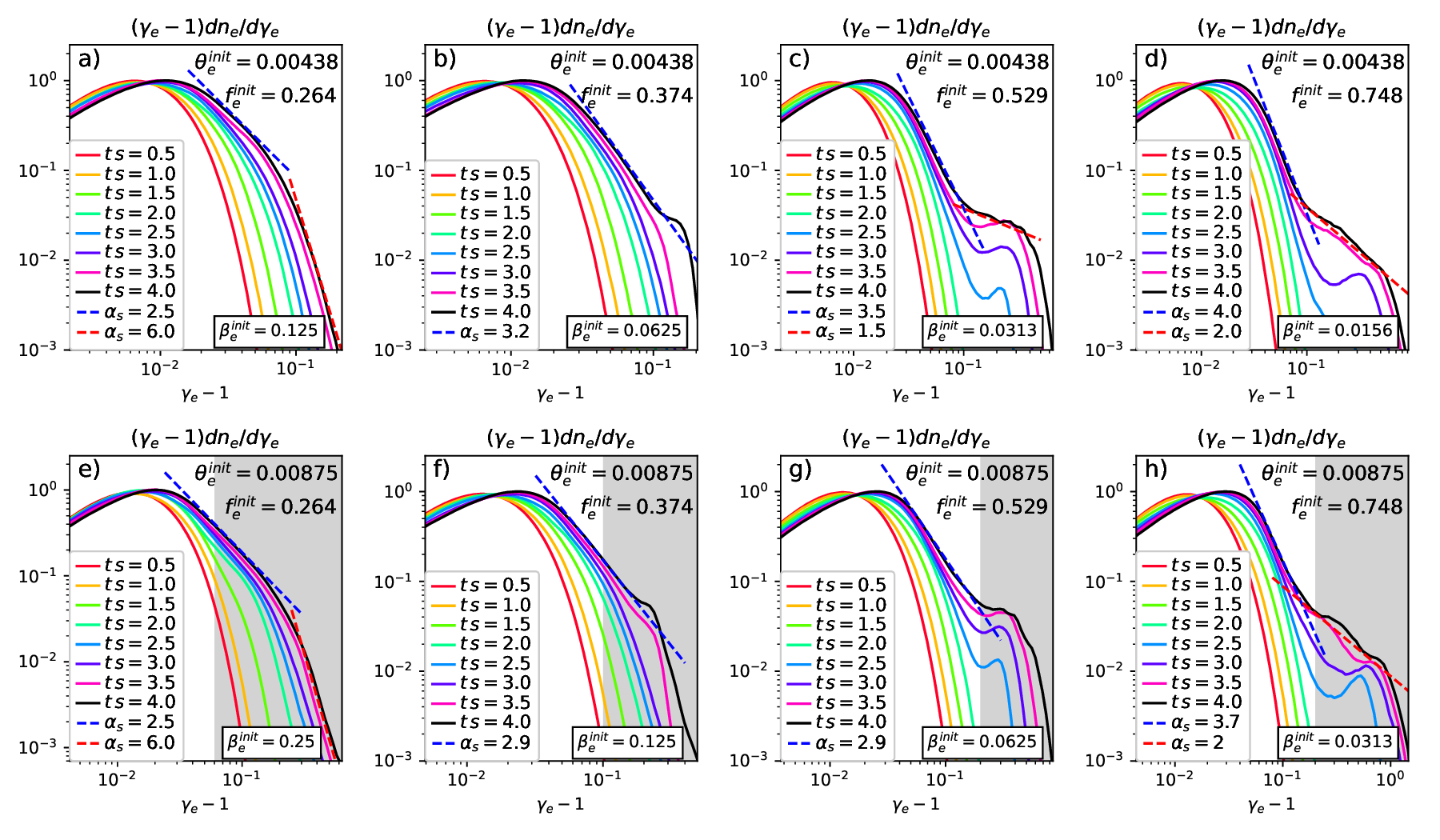}
\caption{The time evolution of the electron spectra for simulations with two values of initial electron temperature, $\Theta_e^{\textrm{init}} = 0.00438$ and $0.00875$ (upper and lower row), and four values of $f_e^{\textrm{init}} = 0.264$, $0.374$, $0.529$, and $0.748$. All runs use $\omega_{c,e}^{\textrm{init}}/s=1200$. Panels $a$-$h$ correspond to runs Fe1Te1-1200, Fe2Te1-1200, Fe3Te1-1200, Fe4Te1-1200, Fe1Te2-1200, Fe2Te2-1200, Fe3Te2-1200, and Fe4Te2-1200 from Table \ref{tab:param}.}
\label{fig:spectra} 
\end{figure*}
A 2D snapshot of the unstable modes shortly after the saturation of $\delta \textbf{\textit{B}}$ (at $t \cdot s = 1.6$) is presented in Figure~\ref{fig:2dstructure}$a$, which displays the $z$-component of the magnetic fluctuations, $\delta \textit{B}_z$. This panel reveals that the dominant fluctuations at this stage are associated with modes with oblique wavevectors relative to the mean magnetic field $\langle \textbf{\textit{B}} \rangle$. The direction of $\langle \textbf{\textit{B}} \rangle$ is indicated by the black arrows. The dominance of oblique modes is, however, transient. By $t \cdot s = 3.1$, shown in Figure~\ref{fig:2dstructure}$b$, the fluctuations propagate predominantly along the direction of $\langle \textbf{\textit{B}} \rangle$, indicating that quasi-parallel modes have become dominant. 

To quantify and evaluate the transition from oblique to quasi-parallel dominance, we compute the magnetic energy associated with each mode type. We categorize the modes as ``oblique'' or ``quasi-parallel'' depending on whether the angle between their wavevector $\textbf{\textit{k}}$ and the mean field $\langle \textbf{\textit{B}} \rangle$ is greater or less than $20^\circ$, respectively. The time evolution of the magnetic energy in each mode category is shown in Figure~\ref{fig:mubreak}$a$, in which the solid- and dashed-black lines represent the magnetic energies in the oblique ($\langle \delta B_{ob}^2 \rangle$) and quasi-parallel ($\langle \delta B_{qp}^2 \rangle$) components, respectively. This analysis reveals that oblique modes dominate until $t \cdot s \approx 2$, after which quasi-parallel modes experience a rapid growth, taking over as the main contributors to the total magnetic fluctuation energy at $t \cdot s \approx 2.2$. This transition is consistent with the change in the evolution of $\Theta_{e,\perp}$ and  $\Theta_{e,\parallel}$ observed at $t \cdot s \approx 2$ in Fig. \ref{fig:mubreak}$b$, as well as with the change in the overall orientation of the dominant modes observed in Figures~\ref{fig:2dstructure}$a$ and \ref{fig:2dstructure}$b$. 

As discussed in Paper ~I, a distinctive feature of the oblique modes is their dominant electrostatic electric field. This can be seen in Fig. \ref{fig:mubreak}$c$, which shows the evolution of the energy in electric field fluctuations, separating it into electrostatic and electromagnetic components, which are shown in solid- and dashed-blue lines, respectively.\footnote{This is done by separating the Fourier-transformed electric field $\delta \tilde{\textbf{\textit{E}}}$ into parts satisfying $\textbf{\textit{k}} \times \delta \tilde{\textbf{\textit{E}}} = 0$ (electrostatic) and $\textbf{\textit{k}} \cdot \delta \tilde{\textbf{\textit{E}}} = 0$ (electromagnetic).} At $t \cdot s \approx 2.2$, the electromagnetic electric fluctuations initiate a rapid growth, becoming dominant over the electrostatic fluctuations at $t \cdot s \approx 2.7$. 

The observed transition from oblique modes characterized by primarily electrostatic fields to quasi-parallel modes characterized by primarily electromagnetic fields aligns well with predictions from linear Vlasov--Maxwell theory. As shown in Paper ~I and using the NHDS solver developed by \cite{2018RNAAS...2...13V}, a transition from oblique quasi-electrostatic (OQES) to parallel electromagnetic z (PEMZ) modes is expected  at $f_e \sim 1.2$. As we can see from the upper horizontal axes in the three panels of Fig. ~\ref{fig:mubreak}, which show the instantaneous value of $f_e$ within the simulation calculated with the average magnetic field and electron density, at the time of the transition ($t \cdot s \approx 2.2$), $f_e \approx 1.2$, which is in good agreement with the expectations from linear theory.\footnote{Our estimate of a transition at $f_e \sim 1.2$ from linear theory in Paper~I assumes $\Theta_{e,\parallel}=0.06$ and $\gamma_g/\omega_{ce} = 10^{-2}$, where $\gamma_g$ is the maximum growth rate of the competing modes. We see in Fig. \ref{fig:mubreak}$b$ that $\Theta_{e,\parallel}\approx0.06$ at the time of the transition ($t \cdot s \approx 2.2$), while  Fig. \ref{fig:mubreak}$a$ shows that $\gamma_g \approx 10\,s$ for the exponential growth phase of both oblique and quasiparallel modes. Given that run Fe3Te2-1200 has $\omega_{ce}^{\textrm{init}}/s=1200$, the modes under consideration have $\gamma_g/\omega_{c,e} \approx 10^{-2}$, making the linear Vlasov--Maxwell theory calculations from Paper I quite suitable for comparison with our simulation results.}

\section{Nonthermal Electron Acceleration under Varying Initial Conditions}
\label{sec:enta}

\noindent Paper~I demonstrates that under the initial conditions of our fiducial run Fe3Te2-1200 ($\Theta_e^{\textrm{init}} = 0.00875$ and $f_e = 0.529$), electrons are efficiently accelerated through inelastic scattering by unstable plasma modes. By the end of the simulation, the resulting electron energy spectra develop a nonthermal tail that resembles a power law with spectral index $\alpha_s$ defined by
\begin{equation}
    \frac{dn_e}{d\gamma_e} \propto (\gamma_e-1)^{-\alpha_s},
\end{equation}
reaching $\alpha_s \sim 2.9$, along with a high-energy bump-like break at $\sim$150 keV ($\gamma_e$ is the electron Lorentz factor). In this section, we investigate how changing the initial plasma parameters modifies the time evolution of the nonthermal electron spectra. 

\subsection{Time evolution of the spectra}
\label{sec:timeevol} 
Figure \ref{fig:spectra} shows the time evolution of the electron spectra in simulations with two values of initial electron temperature, $\Theta_e^{\textrm{init}} = 0.00438$ and $0.00875$, and four initial values for the $f_e$ parameter, $f_e^{\textrm{init}} = 0.264$, $0.374$, $0.529$, and $0.748$ (all of them with $\omega_{ce}^{\textrm{init}}/s=1200$). By comparing its first and second rows, corresponding to temperatures $\Theta_e^{\textrm{init}} = 0.00438$ and 0.00875, we find that the value of the initial temperature in the applied range has only a modest effect on the spectral shape and evolution. In contrast, when comparing the different columns from left to right corresponding to $f_e^{\textrm{init}} = 0.264$, $0.374$, $0.529$, we see that rather modest increases in $f_e^{\textrm{init}}$ (by a factor $\sqrt{2}$ for consecutive cases) leads to a significant change in the shape of the nonthermal tail of the electron spectrum. In particular, at the end of the simulations with the smallest $f_e^{\textrm{init}}= 0.264$, the spectra are characterized by double power-laws with downward spectral breaks near $\gamma_e -1 \approx 0.1-0.2$ (equivalent to $\varepsilon_b \sim$ 50-100 keV). The power-laws have spectral indices with values $\alpha_s \approx 2.5$ and $\approx 6$ for energies below and above the break, respectively. By the end of the simulations with the highest $f_e^{\textrm{init}} = 0.748$, the spectra are significantly harder and characterized by upward spectral breaks at similar energies. In this case, the spectral indices are $\alpha_s \approx 3.7-4$ below the break, and $\approx 2$ above the break. The high-energy parts of the spectra also contain some bumpiness, especially in the case with temperature $\Theta_e^{\textrm{init}} = 0.00875$. Finally, the simulations with the intermediate values of $f_e^{\textrm{init}} =$ 0.374 and 0.529 tend to give single power-laws, which, by the end of the simulations, produce a high-energy bump-like break at $100$-$150\,$keV. 

These results depend only weakly on $T_e^{\mathrm{init}}$ in the shown range of parameters, so $f_e^{\textrm{init}}$ is indeed the crucial parameter for determining the spectral evolution in our simulations. As we further explain in \S \ref{sec:transition} and \ref{sec:works}, this occurs essentially because an increase in $f_e^{\textrm{init}}$ causes a transition in the electron acceleration  from being dominated by OQES modes to being dominated by PEMZ modes, with each of these modes imprinting different features in the electron spectrum.
\begin{figure*}[t!]  
\centering 
\hspace*{-.2cm}\includegraphics[width=18.5cm]{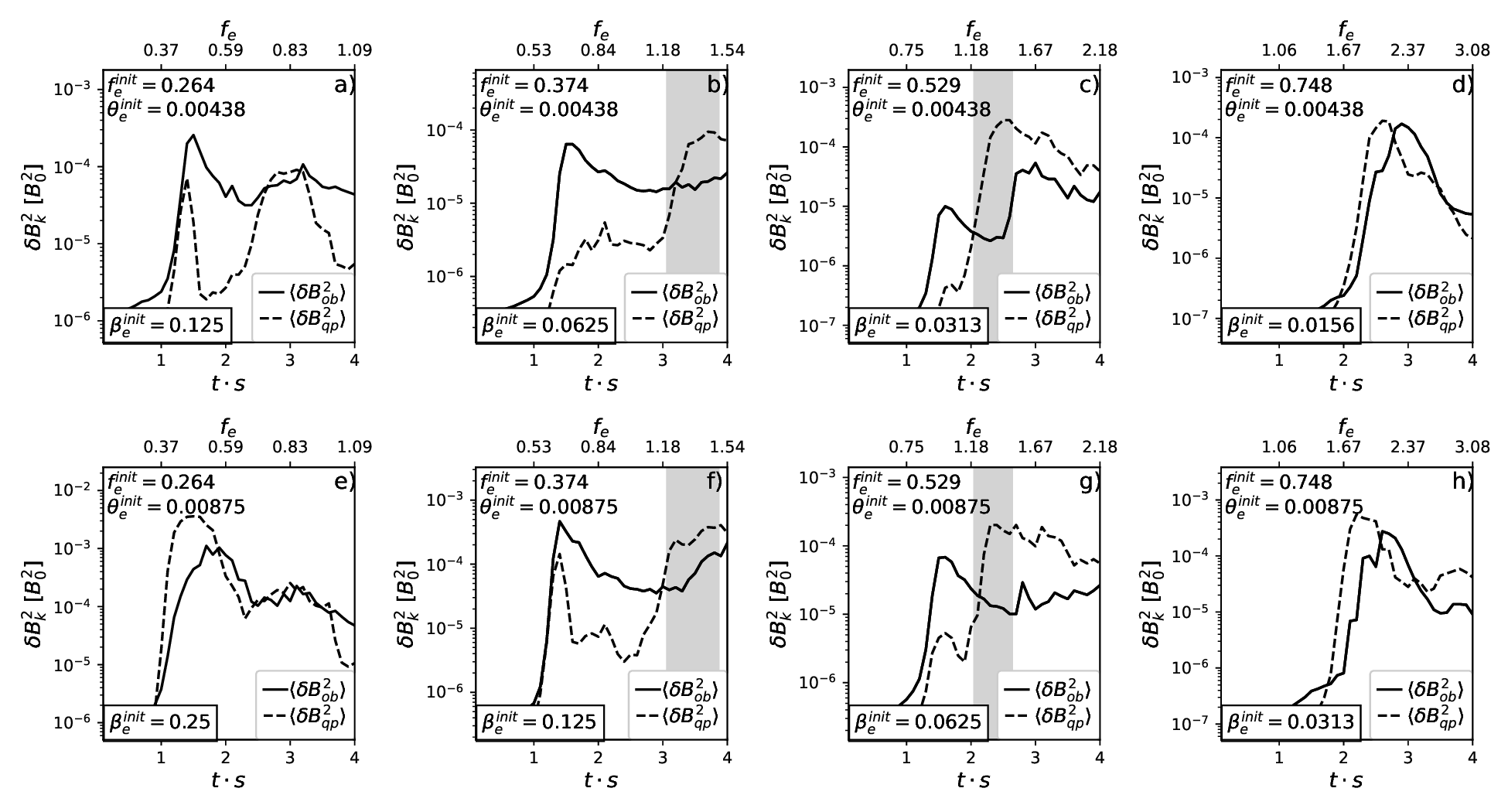}
\caption{The evolution of the energy in magnetic fluctuations for the same runs shown in Fig. \ref{fig:spectra} separated by oblique and quasi-parallel modes. The magnetic energies in these two types of modes ($\delta B_{ob}^2$ and $\delta B_{qp}^2$) are shown in solid-black and dashed-black lines, respectively.} 
\label{fig:compdelb} 
\end{figure*}
\begin{figure*}[t!]  
\centering 
\hspace*{-.2cm}\includegraphics[width=18.5cm]{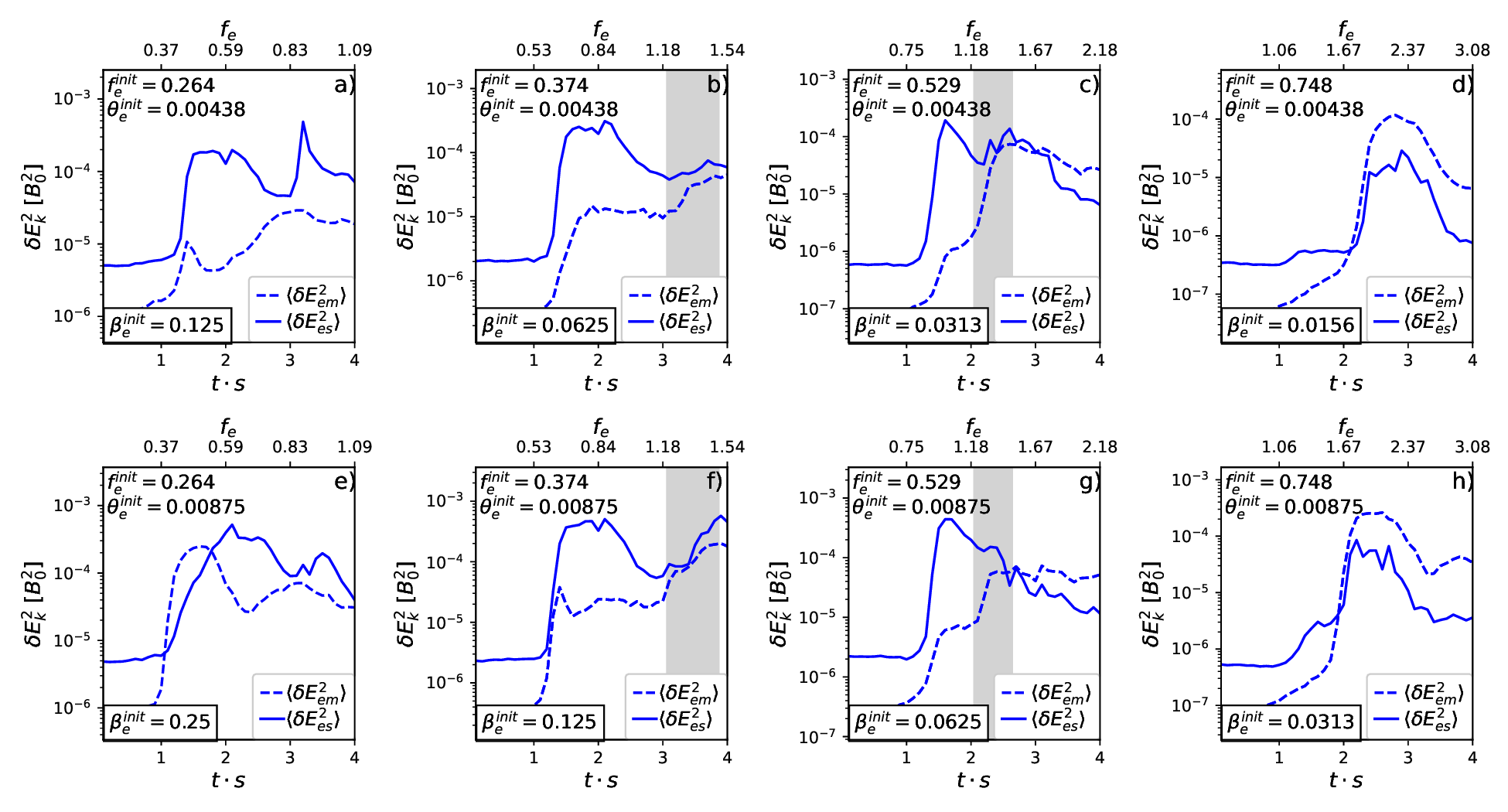}
\caption{The evolution of the energy in electric field fluctuations for the same runs shown in Figs. \ref{fig:spectra} and \ref{fig:compdelb} separated by electrostatic and electromagnetic components. The electric field energies in these two types of modes ($\delta E_{\textrm{es}}^2$ and $\delta E_{\textrm{em}}^2$) are shown in solid-blue and dashed-blue lines, respectively.} 
\label{fig:compdele} 
\end{figure*}
%extend the duration over which PEMZ modes dominate, while simultaneously reducing the time span of OQES mode dominance. Given that in the fiducial case Fe3Te2-1200 with $\Theta_e^{\textrm{init}} = 0.00875$ and $f_e = 0.529$ most of the electron acceleration is attributed to PEMZ modes (as shown in Paper~I), it is natural that an increased prevalence of PEMZ modes caused by increasing $f_e^{\textrm{init}}$ leads to harder spectra. 

\subsection{OQES - PEMZ transition}
\label{sec:transition}

Given the fairly temperature-independent transition from the dominance of OQES to PEMZ modes at $f_e \sim 1.2$, we expect to observe a growing dominance of PEMZ over OQES modes with increasing $f_e^{\textrm{init}}$. This is indeed what we see in Fig. \ref{fig:compdelb}, which shows the time evolution of the energy in the fluctuating magnetic field for the same simulations shown in Fig. \ref{fig:spectra}, decomposing the fluctuations into oblique ($\delta B^2_{\textrm{ob}}$) and quasi-parallel ($\delta B^2_{\textrm{qp}}$) modes, as defined in \S \ref{sec:nature}. Figs. \ref{fig:compdelb}$a$-\ref{fig:compdelb}$d$ correspond to the runs with $\Theta_e^{\textrm{init}} = 0.00438$ and $f_e^{\textrm{init}} = 0.264$, $0.374$, $0.529$, and $0.748$ (the same simulations of which the electron spectral evolution is shown in Figs. \ref{fig:spectra}$a$-\ref{fig:spectra}$d$). In the lowest $f_e$ case ($f_e^{\textrm{init}} = 0.264$, Fig. \ref{fig:compdelb}$a$), $\delta B^2_{\textrm{ob}} \gtrsim \delta B^2_{\textrm{qp}}$ essentially throughout the whole simulation. As $f_e^{\textrm{init}}$ increases to 0.374 (Fig. \ref{fig:compdelb}$b$) and 0.529 (Fig. \ref{fig:compdelb}$c$), quasi-parallel modes experience a rapid exponential growth when $f_e \sim 1.2$ (the instantaneous $f_e$ is plotted in the upper horizontal axes in all the panels of Fig. \ref{fig:compdelb}). After that, the quasi-parallel modes reach saturation around $f_e \sim 1.5$ (we mark the $f_e=1.2-1.5$ interval in Figs. \ref{fig:compdelb}$b$ and \ref{fig:compdelb}$c$ using a grey background). For the highest $f_e^{\textrm{init}} = 0.748$ (Fig. \ref{fig:compdelb}$d$), quasi-parallel modes dominate from the onset of magnetic fluctuation growth, which occurs when $f_e \sim 1.7$,  although oblique modes become prominent again at later times. Overall, this behavior is consistent with the dominance of PEMZ modes at $f_e \gtrsim 1.2-1.5$, which suggests that these modes are more relevant for electron acceleration as $f_e^{\textrm{init}}$ increases. %We notice that the observed transition at $f_e \sim 1.2$ for the runs with $\Theta_e^{\textrm{init}} = 0.00438$ is essentially the same that happens for our fiducial run Fe3Te2-1200, which uses $\Theta_e^{\textrm{init}} = 0.00875$. This reflects a weak dependence of this transition on temperature, consistent with what we obtained from linear Vlasov theory in Paper~I. 

Since OQES modes are primarily associated with electrostatic electric fields, this type of electric field is expected to dominate in the case with the lowest $f_e^{\textrm{init}}$ = 0.264, with the electromagnetic component of the electric field becoming progressively more significant for the runs with larger $f_e^{\textrm{init}}$. This expectation is confirmed in Fig. \ref{fig:compdele}, which shows the evolution of the energies in the electric field fluctuations $\delta \textbf{\textit{E}}$ for the same simulations shown in Figs. \ref{fig:spectra} and \ref{fig:compdelb}, decomposing the electric field energy into electrostatic (solid-blue) and electromagnetic (dashed-blue) components. Focusing first on runs with $\Theta_e^{\textrm{init}}=0.00438$ (first row), we see that for $f_e^{\textrm{init}} = 0.264$ (Fig. \ref{fig:compdele}$a$), the electric field energy is dominated by the electrostatic component, which, combined with knowledge that the modes are mainly oblique (Fig. \ref{fig:compdelb}$a$), supports the expectation that the run with $f_e^{\textrm{init}} = 0.264$ is  dominated by OQES modes. 

For $f_e^{\textrm{init}} = 0.374$ (Fig. \ref{fig:compdele}$b$) and $f_e^{\textrm{init}} = 0.529$ (Fig. \ref{fig:compdele}$c$), the electromagnetic electric field fluctuations experience  rapid growth at $f_e \sim 1.2$, reaching values comparable to the electrostatic field fluctuations at $f_e \sim 1.5$ (this transition is marked by the grey background in Figs. \ref{fig:compdele}$b$ and \ref{fig:compdele}$c$). This behavior occurs at the transition time when the modes become dominantly quasi-parallel, as we show in Figs. \ref{fig:compdelb}$b$ and \ref{fig:compdelb}$c$, supporting the transition from OQES-dominated to PEMZ-dominated modes during the $f_e \sim 1.2-1.5$ period. 

Finally, in the case $f_e^{\textrm{init}} = 0.748$ (Fig. \ref{fig:compdele}$d$), the electric field energy is dominated by its electromagnetic component right from the moment when the instabilities are triggered, which is consistent with the dominance of PEMZ modes.
\begin{figure*}[t!]  
\centering 
\hspace*{0cm}\includegraphics[width=18.3cm]{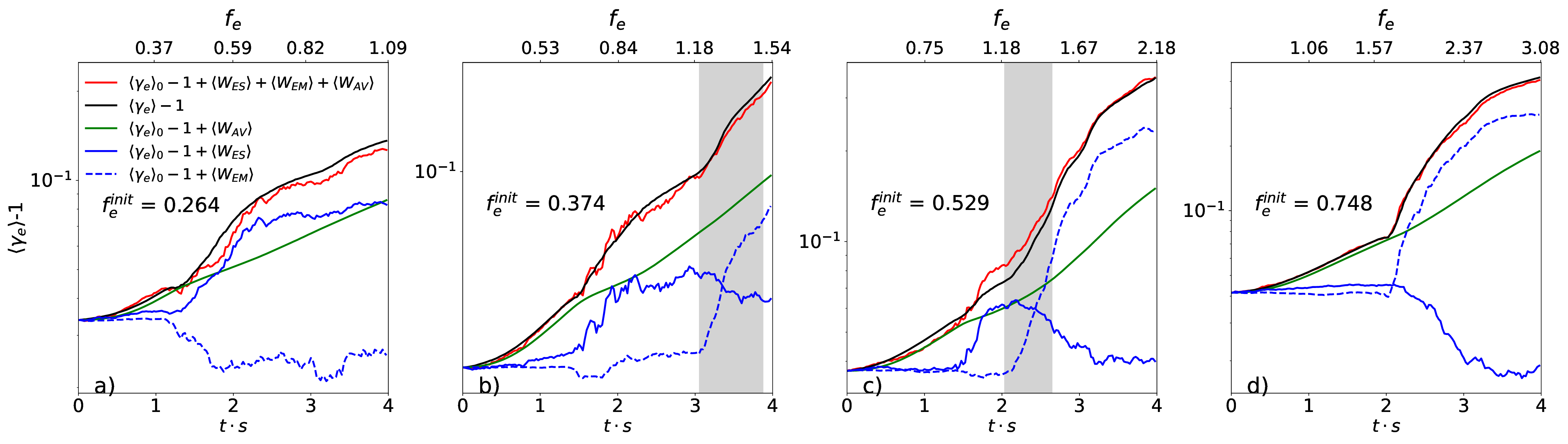}
\caption{The evolution of the average energy normalized to $m_ec^2$, $\langle \gamma_e \rangle -1$, for the highest-energy electrons in our simulations with $\Theta_e^{\textrm{init}} = 0.00875$ and $f_e^{\textrm{init}} = 0.264$ (run Fe1Te2-1200; panel $a$), 0.374 (run Fe2Te2-1200; panel $b$), 0.529 (run Fe3Te2-1200; panel $c$), and 0.748 (run Fe4Te2-1200; panel $d$). The analyzed electrons in each case have by the end of the simulations ($t\cdot s = 4$) energies $\gamma_e -1$ larger than 0.06, 0.1, 0.2 and 0.2, respectively (these energies are marked by the grey backgrounds in the second row of Fig. \ref{fig:spectra}). In each panel, the average electron energies are decomposed into three contributions: (i) the work done by electrostatic electric fields ($\langle W_{es} \rangle$, solid-blue lines), (ii) the work done by electromagnetic electric fields ($\langle W_{em} \rangle$, dashed-blue lines), and (iii) the energization by anisotropic viscosity ($\langle W_{AV} \rangle$, green line). For each of these energies, we add the initial average energy ($\langle \gamma_e \rangle_0 -1$). The red line shows $\langle \gamma_e \rangle_0 -1 + \langle W_{es} \rangle + \langle W_{em} \rangle + \langle W_{AV} \rangle$, which we compare with the actual average energy measured directly from the simulation, $\langle \gamma_e \rangle - 1$, shown in black.} 
\label{fig:works} 
\end{figure*}

Although the analysis above focuses on simulations with an initial temperature $\Theta_e^{\textrm{init}} = 0.00438$, runs with $\Theta_e^{\textrm{init}} = 0.00875$ exhibit a very similar behavior (as already shown for run Fe3Te2-1200 in \S \ref{sec:nature}). This similarity is especially true for the runs with $f_e^{\textrm{init}} = 0.374$, $0.529$, and $0.748$, for which the evolution of the electric and magnetic fluctuations show almost no difference between the cases with $\Theta_e^{\textrm{init}} = 0.00438$ and $0.00875$, as shown by the first and second rows of Figs. \ref{fig:compdelb} and \ref{fig:compdele}. The only noticeable difference arises for $f_e^{\textrm{init}} = 0.264$, in which case the runs with $\Theta_e^{\textrm{init}} = 0.00875$ show a prevalence of quasi-parallel electromagnetic modes at $f_e \lesssim 0.6$ that does not appear for $\Theta_e^{\textrm{init}} = 0.00438$ when comparing the obliquity of the modes in Figs. \ref{fig:compdelb}$a$ and \ref{fig:compdelb}$e$, as well as the prevalence of electrostatic and electromagnetic modes in Figs. \ref{fig:compdele}$a$ and \ref{fig:compdele}$e$. This behavior is broadly consistent with the predictions from linear Vlasov--Maxwell theory found in Paper~I, which indicate that if $\Theta_{e,\parallel}$ is increased from $\Theta_{e,\parallel} = 0.002$ to 0.006, the minimum value $f_e$ required for the dominance of OQES modes changes from $f_e \sim 0.4$ to $\sim 0.6$, narrowing down the interval of values of $f_e$ in which OQES modes dominate. For values of $f_e$ below these lower limits, parallel electromagnetic whistler waves (PEMW) should dominate. However, despite this initial dominance of PEMW modes, in \S \ref{sec:works}, we show that in the case of $f_e^{\textrm{init}}=0.264$ with $\Theta_e^{\textrm{init}} = 0.00875$, the electron acceleration is entirely dominated by the electrostatic fields that characterize OQES modes. 
%Besides this difference caused by the change in temperature, our results demonstrate that under typical ALT conditions, electron acceleration is mainly sensitive to $f_e^{\textrm{init}}$. Basically, by increasing this parameter from 0.264 to 0.748, we increase the time period in which the PEMZ modes outcompete the OQES modes. When that happens, the acceleration of electrons is expected to be dominated by electromagnetic fields, while when $f_e \lesssim 1.2$, high energy electrons are expected to be energized mainly by electrostatic fields. This is demonstrated below by separating the contribution of electrostatic and electromagnetic fields to electron acceleration for simulations with different $f_e^{\textrm{init}}$.

\subsection{Contributions from electrostatic and electromagnetic energy}
\label{sec:works}

In \S \ref{sec:transition}, we identify the unstable electromagnetic modes that dominate at different times for all of our explored combinations of $\Theta_e^{\textrm{init}}$ and $f_e^{\textrm{init}}$. In this section, we quantify the efficiency with which each of these modes contribute to electron acceleration. In Fig.~\ref{fig:works}, we show the evolution of the average energy gain of the highest-energy electrons in simulations with $\Theta_e^{\textrm{init}} = 0.00875$ and $f_e^{\textrm{init}} = 0.264$, 0.374, 0.529, and 0.748 (Fig.~\ref{fig:works}$a$-$d$). We decompose the electron energy gains into three contributions. The first and second contributions correspond to the works done by the electrostatic and electromagnetic electric fields on each electron ($W_{es}$ and $W_{em}$), which we normalize by $m_ec^2$ and calculate as
\begin{equation}
    W_k(t) = \int_{0}^t dt\,\textbf{\textit{v}}\cdot (-e)\delta \textbf{\textit{E}}_k/m_ec^2,
\end{equation}
where $W_k$ represents the works done either by the electrostatic or electromagnetic electric field fluctuations, $\delta \textbf{\textit{E}}_k$ ($k$ symbolizes either ``$\it{es}$'' or ``$\it{em}$'', depending on the electric field being electrostatic or electromagnetic), $\textbf{\textit{v}}$ is the electron's velocity and $-e$ is the electron's electric charge. The values of these contributions averaged over all the electrons, $\langle W_{es} \rangle$ and $\langle W_{em} \rangle$, are shown in Fig. \ref{fig:works}$a-d$ in solid-blue and dashed-blue lines, respectively. The third contribution is the energization by the so-called ``anisotropic viscosity'' ($W_{AV}$; whose average value $\langle W_{AV} \rangle$ is shown by the green lines in Fig. \ref{fig:works}$a-d$). This contribution arises from the electron pressure anisotropy with respect to the local magnetic field, through which the electrons tap into the bulk shear motion of the plasma, gaining energy at a rate \citep{1983bpp..conf....1K,1997PhPl....4.3974S}:
\begin{equation}
\frac{dU_{e}}{dt} = r \, \Delta p_e,
\label{eq:av}
\end{equation}
where $U_e$ is the electron kinetic energy density, $r$ is the magnetic field growth rate (in our setup, $r = -sB_xB_y/B^2$), and $\Delta p_e = p_{e,\perp} - p_{e,\parallel}$ denotes the pressure anisotropy based on the pressures perpendicular ($p_{e,\perp}$) and parallel ($p_{e,\parallel}$) to the magnetic field.\footnote{In Paper~I, we show that anisotropic viscosity accounts for essentially all of the total electron heating in our shearing setup.} 

Figure~\ref{fig:works}$a$ shows the case $f_e^{\textrm{init}} = 0.264$ (run Fe1Te2-1200). We analyze 2700 electrons that have reached $\gamma_e - 1 > 0.06$ by the end of the simulation. This population, representative of the nonthermal tail, is marked by the grey background in Fig.~\ref{fig:spectra}$e$. In this case, the total energy gain from electric fields is dominated by the electrostatic component during the whole simulation, showing the dominant role of OQES modes in the acceleration process. The OQES modes are dominant despite the strong initial growth of PEMW modes. In fact, these modes appear to gain energy from the highest-energy electrons, based on their negative contribution to electron acceleration during the time at which they dominate ($t\cdot s \sim 1-1.8$). The fastest energization by electrostatic fields occurs in the interval $t\cdot s \sim 1.3$–$2$, coinciding with the rapid growth and saturation of oblique electrostatic modes, as can be seen from Figs.~\ref{fig:compdelb}$e$ and \ref{fig:compdele}$e$. After this transition, electrostatic fluctuations continue to energize the electrons but at a much lower rate.

We find similar behavior for the runs with $f_e^{\textrm{init}} = 0.374$ and $0.529$. We analyze the energy gain of 3200 and 2700 electrons that have reached $\gamma_e - 1 > 0.1$ and 0.2 by the end of their respective simulations (these electron populations are marked by the grey backgrounds in Figs.~\ref{fig:spectra}$f$ and \ref{fig:spectra}$g$). In both simulations, the electrostatic electric field components initially dominates the energy gain, producing a rapid energy increase that approximately coincides with the growth and saturation of the OQES modes, as can be seen in Figs.~\ref{fig:compdelb}$f$–$g$ and \ref{fig:compdele}$f$–$g$. As in the case with $f_e^{\textrm{init}} = 0.264$, this stage is followed by a much slower increase in electric field energization, which is still led by the electrostatic field and lasts until $f_e \sim 1.2$. Beyond this time, the electromagnetic fluctuations take over as the main energization channel. For $f_e^{\textrm{init}} = 0.529$, the electromagnetic contribution grows exponentially until $f_e \sim 1.5$. Thus, the fastest acceleration in this run occurs in the interval $f_e \sim 1.2$–$1.5$, coinciding with the transition from OQES to PEMZ mode dominance, marked by the grey background shade in Figs.~\ref{fig:compdelb}$g$ and \ref{fig:compdele}$g$. Afterwards, the electromagnetic fields continue to energize the electrons, but at a progressively lower rate, which approaches zero by the end of the simulation. In the case of $f_e^{\textrm{init}} = 0.374$, there is also a period of nearly exponential growth of the energy gain by the electromagnetic electric field which lasts until the end of the simulation ($f_e \sim 1.5$). That period coincides with the transition from OQES to PEMZ mode dominance, also marked with a grey background in Figs.~\ref{fig:compdelb}$f$. 

Finally, for $f_e^{\textrm{init}} = 0.748$, we analyze the energization for 3000 electrons that reach $\gamma_e - 1 > 0.2$ by the end of the simulation, which are marked by the grey background in Fig.~\ref{fig:spectra}$h$. In this regime, the electromagnetic contribution remains consistently positive and dominant throughout the whole run, while the electrostatic work is negative. This behavior is consistent with the expected prevalence of acceleration by PEMZ modes. As in the case with $f_e^{\textrm{init}} = 0.529$, the energy gain by electric fields gradually decreases, eventually approaching zero when $f_e \gtrsim 2.4$.

These results suggest that the fastest stages of electron acceleration, either by OQES or PEMZ modes, occur during time periods at which these modes undergo exponential growth, and last until they reach saturation. In the run in which the OQES-PEMZ transition is fully captured (i.e., run Fe3Te2-1200, with $\Theta_e=0.00875$ and $f_e^{\textrm{init}} =$ 0.529), this period of rapid acceleration occurs in the well defined range $f_e \sim 1.2-1.5$. In addition, the energy gain given by anisotropic viscosity is in all cases comparable to the one of the electric field, only becoming slightly subdominant as $f_e^{\textrm{init}}$ increases. The four panels in Fig. \ref{fig:works} also show the total energy evolution of the respective electron populations with a black line, and in red the sum of the works done by the anisotropic viscosity and by the electrostatic and electromagnetic components of the electric field. This sum agrees well with the total energy of the electrons, which is consistent with the fact that the electron energy evolution is well accounted for by these three contributions.\footnote{The largest discrepancy appears in the case of $f_e^{\textrm{init}}=0.264$, which exhibits a $10\%$ difference between the red and black lines. This discrepancy can be explained by the fact that we calculated the work done by the electrostatic and electromagnetic electric fields using a time step that is about $3000$ times greater than the one used in the simulations.}  

\begin{figure*}[t!]  
\centering 
\hspace*{-.2cm}\includegraphics[width=18.5cm]{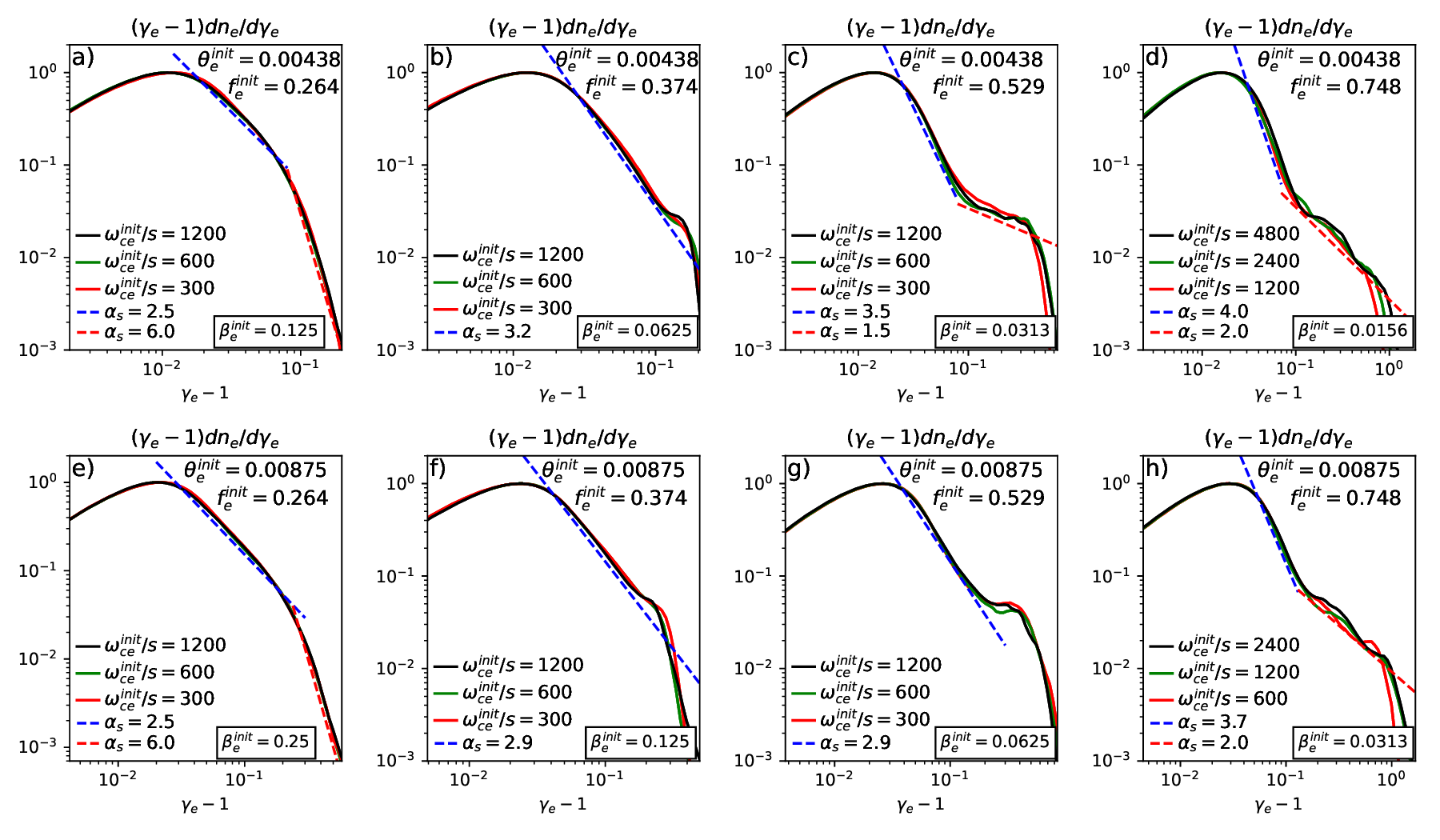}
\caption{The final electron spectra (at $t \cdot s = 4$) from runs with identical plasma parameters, but with $\omega_{ce}^{\textrm{init}}/s$ differing by a factor of 4 between the smallest and largest values (the runs with $\omega_{ce}^{\textrm{init}}/s=1200$ are the same ones as shown in Figure~\ref{fig:spectra}). No significant differences are observed in the final spectra, indicating that $\omega_{ce}^{\textrm{init}}/s$ does not substantially influence the efficiency of nonthermal electron acceleration over the applied range of parameters.} 
\label{fig:compmag} 
\end{figure*} 

\subsection{Extrapolation to the (realistic) very high $\omega_{ce}/s$ regime}
\label{sec:realistic}

As discussed in \S \ref{sec:setup}, realistic solar flare conditions involve values of $\omega_{ce}/s$ of the order of $10^9-10^{10}$, far beyond the values achievable in our simulations. This raises two key questions: (1) Do PEMW, PEMZ, and OQES modes remain dominant in the same plasma regimes when extrapolated to realistic $\omega_{ce}/s$ values? (2) If so, does their impact on electron acceleration remain unchanged for different $\omega_{ce}/s$? 

To address these questions, we provide similar arguments as in Paper I. For the first question, we note that in our shearing setup the growth rate $\gamma_g$ of the unstable modes is expected to scale with $s$, since $s^{-1}$ sets the timescale for the evolution of the macroscopic plasma conditions. Therefore, modes that regulate the electron temperature anisotropy during the various stages of the simulation must grow on a timescale comparable to $s^{-1}$ to become dynamically relevant. Therefore, in the realistic regime with large $\omega_{ce}^{\textrm{init}}/s$, the relevant instabilities are those with thresholds corresponding to comparatively large values of $\omega_{ce}/\gamma_g$. 

In Paper~I, we employ linear Vlasov--Maxwell theory to identify the dominant instabilities across our range of $f_e$ and $\Theta_{e,\parallel}$ values. By exploring $\gamma_g/\omega_{ce} = 10^{-6}$ and $10^{-2}$, we find that the dominant modes responsible for the regulation of temperature anisotropy do not change significantly between these cases. This strongly suggests that the regime boundaries defined by mode dominance in our simulations can be extrapolated to realistic $\omega_{ce}/s$ values.\footnote{In each regime, the dominant instability is identified as the one requiring the smallest anisotropy threshold under given plasma conditions and for a fixed value of $\omega_{ce}/\gamma_g$.} 

To address the second question, whether realistic $\omega_{ce}^{\textrm{init}}/s$ values affect the spectral evolution of electrons, we compare in Figure~\ref{fig:compmag} the final electron spectra (at $t \cdot s = 4$) from runs with identical plasma parameters as in Figure~\ref{fig:spectra}, but with $\omega_{ce}^{\textrm{init}}/s$ differing by a factor of 4. We find no significant differences in the final spectra across these cases, indicating that $\omega_{ce}^{\textrm{init}}/s$ does not substantially influence the efficiency of nonthermal electron acceleration in this parameter range. 

We interpret this observed independence of the acceleration process of $\omega_{ce}^{\textrm{init}}/s$ by considering the relationship between the effective pitch-angle scattering rate due to the instabilities, $\nu_{\textrm{eff}}$, and the shear rate $s$. As discussed in detail in Paper~I, $\nu_{\textrm{eff}}$ is expected to scale proportionally to $s$, since $s$ drives the growth of anisotropies, while $\nu_{\textrm{eff}}$ governs their suppression. Given that the anisotropies evolve toward a marginally unstable, quasi-steady state, we expect $\nu_{\textrm{eff}} \propto s$. Because of this scaling, the average number of effective pitch-angle deflections experienced by electrons over a fixed interval of the normalized time $t \cdot s$ should remain largely independent of $\omega_{ce}^{\textrm{init}}/s$. Consequently, in a stochastic acceleration scenario, the acceleration efficiency over a fixed number of shear timescales should primarily depend on the dispersive properties of the unstable modes \citep[see, e.g.,][]{1998JGR...10320487S}, which are not expected to vary significantly with $\omega_{ce}^{\textrm{init}}/s$. In fact, as long as $\omega_{ce}^{\textrm{init}} \gg s$, the unstable modes oscillate and propagate on short timescales ($\sim \omega_{ce}^{-1}$), effectively decoupled from the slowly varying background dynamics ($\sim s^{-1}$). Altogether, these arguments support our conclusion that the efficiency of electron acceleration observed in our simulations should remain valid in realistically high $\omega_{ce}^{\textrm{init}}/s$ regimes. 

\section{Summary and conclusions}
\label{sec:conclusions}

In this work, we investigate the efficiency of electron acceleration by temperature anisotropy instabilities under plasma conditions representative of above-the-loop-top (ALT) regions of solar flares. In order to capture the long-term effect of the instabilities on the electron spectra, we use 2D fully kinetic PIC simulations of a shearing plasma that amplifies the background magnetic field by a factor of $\sim4$, focusing only on the electron-scale dynamics (ions are assumed immobile). This way temperature anisotropies are self-consistently produced due to the adiabatic evolution of the electron temperatures, eventually triggering the instabilities and maintaining them in their saturated, marginally unstable stage for a few shear times ($s^{-1}$), as seen in Fig. \ref{fig:mubreak}$a$. We envision this magnetic amplification process to occur locally within patches of plasma turbulence in ALT regions. 

We explore a parameter space that considers initial electron temperatures of $T_e^{\mathrm{init}} = 26$ and $52$~MK, corresponding to $\Theta_e^{\mathrm{init}} \equiv k_B T_e^{\mathrm{init}} / m_e c^2 = 0.00438$ and $0.00875$ and initial values for the ratio $f_e = \omega_{c,e}/\omega_{p,e}$ of $f_e^{\mathrm{init}} = 0.264$, $0.374$, $0.529$, and $0.748$. 

\subsection{Summary of simulation results}
\label{sec:summary}

We summarize our main findings as follows:

\begin{itemize}\setlength\itemsep{0.4em}
\item \textit{Spectral diversity:} The resulting electron spectra develop nonthermal high-energy tails with double power-law shapes and break energies in the range $\varepsilon_b \sim 50$--$150\,$keV. Runs with $f_e^{\textrm{init}}=0.264$ produce downward (knee-like) breaks (Figs.~\ref{fig:spectra}$a$, \ref{fig:spectra}$e$), whereas runs with $f_e^{\textrm{init}}=0.748$ yield upward (elbow-like) breaks (Figs.~\ref{fig:spectra}$d$, \ref{fig:spectra}$h$). Simulations with intermediate values of $f_e^{\textrm{init}}$ tend to produce single power-laws terminating in a bump-like break around $100$--$150\,$keV. These results depend only weakly on $T_e^{\mathrm{init}}$, indicating that $f_e^{\textrm{init}}$ is the key parameter controlling the spectral evolution for typical ALT temperatures of a few tens of MK.
    
\item \textit{Dominance of different modes:} The strong dependence of the spectral evolution on $f_e^{\textrm{init}}$ is closely linked to the type of unstable mode that dominates the system. When the final spectra exhibit downward breaks (runs with $f_e^{\textrm{init}} = 0.264$), the electron acceleration is dominated by OQES modes. When the final nonthermal tails display upward breaks (runs with $f_e^{\textrm{init}} = 0.748$), the electron acceleration is instead dominated by PEMZ modes. For intermediate values of $f_e^{\textrm{init}}$ ($0.374$ and $0.529$), the unstable modes exhibit a transition from OQES to PEMZ dominance as $f_e$ crosses $1.2$-$1.5$, and both types of modes contribute to the acceleration. Regardless of the dominant mode, the fastest energization always occurs during the exponential growth phase of the instability, ceasing once the modes saturate. In the runs that fully capture the OQES-to-PEMZ transition  (with $f_e^{\textrm{init}} = 0.374$ and 0.529), this rapid-acceleration window coincides with the transition itself ($f_e \approx 1.2$-$1.5$), highlighted by the gray background in Figs.~\ref{fig:works}$b$ and \ref{fig:works}$c$.

\item \textit{Independence of time-scale separation:} Our results show no dependence on the ratio between the initial electron cyclotron frequency $\omega_{ce}^{\textrm{init}}$ and the shear rate $s$, as expected for a stochastic acceleration scenario in which the effective pitch-angle scattering rate scales with $s$. This suggests that our results can be safely applied to ALT environments, where this ratio is several orders of magnitude greater than unity.
\end{itemize}

\subsection{Comparison with observations}
\label{sec:observations}

The dependence of spectral hardness and break type on $f_e$ revealed by our simulations suggests several possible connections to observations of flare-accelerated electrons:

\begin{itemize}\setlength\itemsep{0.4em}
    \item When $f_e \lesssim 1.2$ (as in our OQES-dominated runs with $f_e^{\textrm{init}} = 0.264$), the temperature anisotropy instability produces double power-law spectra with downward (knee-like) breaks. These types of spectra are the most commonly observed, with HXR footpoint spectra typically showing break energies near $\varepsilon_b \sim 100$~keV and spectral indices below and above the break in the ranges $\alpha_{s,b} \sim 2.5$-$4.5$ and $\alpha_{s,a} \sim 3.5$-$5.5$ \citep{2019SoPh..294..105A}. Our runs with $f_e^{\textrm{init}} = 0.264$ (Fe1Te1-1200 and Fe1Te2-1200, with $\Theta_e^{\textrm{init}}=0.00438$ and $0.00875$, respectively) yield spectra with $\varepsilon_b \sim 50$ and $150$~keV, respectively, and $\alpha_{s,b} \sim 2.5$ and $\alpha_{s,a} \sim 6$ (Figs.~\ref{fig:spectra}$a$, \ref{fig:spectra}$e$), roughly consistent with observations. A notable example is the limb-occulted X8.2 flare of 2017 September 10, where emission originated near the reconnection site. \citet{2021ApJ...908L..55C} infer a nonthermal electron distribution with $\alpha_{s,b} \sim 3.6$, which steepens above $\varepsilon_b \sim 160$~keV to $\alpha_{s,a} \sim 6$ and closely resembles the spectra of run Fe1Te2-1200, apart from having a steeper low-energy index.

    \item When $f_e$ evolves from $f_e \lesssim 1.2$ to $f_e \gtrsim 1.5$ (as in our runs with $f_e^{\textrm{init}} = 0.374$ and $0.529$), the instability tends to generate single power-law spectra ending in a bump-like break. Similar spectra are reported for nine solar energetic electron (SEE) events observed by Wind/3DP \citep{2025A&A...699A...2L}, with median indices of $2.52_{-0.25}^{+0.29}$ and bump energies near $59^{+18.1}_{-3.2}$~keV. These are broadly consistent with our runs with $f_e^{\textrm{init}} = 0.374$, which produce $\alpha_s = 2.9$--$3.2$ and bump energies near $70$-$100$~keV (Figs.~\ref{fig:spectra}$b$, \ref{fig:spectra}$f$).

    \item When the instability is triggered after the time step when $f_e \gtrsim 1.5$ (as for our PEMZ-dominated runs with $f_e^{\textrm{init}} = 0.748$), the resulting spectra resemble double power-laws with upward (elbow-like) breaks at $\varepsilon_b \sim 50$--$100$~keV, with $\alpha_{s,b} \sim 3.7$--$4$ and $\alpha_{s,a} \sim 2$. These features are broadly consistent with observations of HXR spectra that harden toward hundreds of keV \citep{1988SoPh..118...49D,1975SoPh...43..415S,1985JPSJ...54.4462Y}, suggesting intrinsic electron spectral hardening. \citet{2013ApJ...763...87A} also find that microwave-inferred electron indices are harder than those from HXRs, implying spectral hardening around hundreds of keV. Their sample shows a typical gap $\alpha_{s,b} - \alpha_{s,a} \sim 1.6$, comparable to the $\sim 1.7$--$2$ gap seen in our $f_e^{\textrm{init}}=0.748$ runs.
\end{itemize}

Although our simulations tend to reproduce some key spectral features reported in these and other observations, each run represents a single, idealized plasma configuration, with the electron spectra measured after the mean magnetic field has been amplified by a factor of $\sim 4$. In real ALT regions, electron distributions are likely shaped by a complex combination of plasma parameters and turbulent conditions, and thus a single run should not be expected to replicate an observed flare spectrum in detail. Moreover, our models do not include electron transport effects, which can significantly alter the inferred spectra and represent an additional source of discrepancy. Nevertheless, our results demonstrate that temperature anisotropy instabilities constitute an efficient stochastic acceleration mechanism and offer a promising framework for explaining the observed diversity and temporal evolution of electron spectra in solar flares.

\begin{acknowledgments}
MR gratefully acknowledges support from the ANID-FONDECYT grant 1191673, as well as from the Center for Astrophysics and Associated Technologies (CATA; ANID Basal grant FB210003). DV is supported by STFC Consolidated Grant ST/W001004/1. The numerical simulations used in this research were performed in the supercomputing infrastructure of the NLHPC (CCSS210001) at the Center for Mathematical Modeling (CMM) of University of Chile. 
\end{acknowledgments}

\bibliography{sample701}{}
\bibliographystyle{aasjournal}



\end{document}